\RequirePackage{lineno} 
\documentclass[aps,prl,nobibnotes,nofootinbib,preprintnumbers,amsmath,amssymb,latexsym,array,enumerate,letter,twocolumn,superscriptaddress]{revtex4}
\usepackage{amssymb}
\usepackage{amsmath}
\usepackage{epsfig}
\usepackage{hyperref}
\usepackage{breakurl}
\usepackage{xcolor}
\makeatletter
\def\simgt{\mathrel{\lower2.5pt\vbox{\lineskip=0pt\baselineskip=0pt
           \hbox{$>$}\hbox{$\sim$}}}}
\def\simlt{\mathrel{\lower2.5pt\vbox{\lineskip=0pt\baselineskip=0pt
           \hbox{$<$}\hbox{$\sim$}}}}
\makeatother

\newcommand{\be}{\begin{equation}}
\newcommand{\ee}{\end{equation}}
\newcommand{\bea}{\begin{eqnarray}}
\newcommand{\eea}{\end{eqnarray}}
\newcommand{\beq}{\begin{eqnarray}}
\newcommand{\eeq}{\end{eqnarray}}

\def\lsim{\mathrel{\rlap{\lower4pt\hbox{\hskip1pt$\sim$}}
     \raise1pt\hbox{$<$}}}         %less than or approx. symbol
\def\gsim{\mathrel{\rlap{\lower4pt\hbox{\hskip1pt$\sim$}}
     \raise1pt\hbox{$>$}}}         %greater than or approx. symbol

\begin{document}
%\linenumbers

\title{Searching for Dark Photon Dark Matter in LIGO O1 Data}

\author{Huai-Ke Guo}
\affiliation{Department of Physics and Astronomy, University of Oklahoma, Norman, OK 73019, USA}

\author{Keith Riles}
\affiliation{Department of Physics, University of Michigan, Ann Arbor, MI 48109}

\author{Feng-Wei Yang}
\email{fwyang@hku.hk}
\affiliation{Department of Physics and Laboratory for Space Research, The University of Hong Kong, PokFuLam, Hong Kong SAR, China}
\affiliation{Department of Physics and Astronomy, University of Utah, Salt Lake City, UT 84112, USA}

\author{Yue Zhao}
\affiliation{Department of Physics and Astronomy, University of Utah, Salt Lake City, UT 84112, USA}

\begin{abstract}{
Dark matter exists in our Universe, but its nature remains mysterious. The remarkable sensitivity of the Laser Interferometer Gravitational-Wave Observatory (LIGO) may be able to solve this mystery. A good dark matter candidate is the ultralight dark photon.   
  Because of its interaction with ordinary matter, it induces displacements on LIGO mirrors that can lead to an observable signal. In a study that bridges gravitational wave science and particle physics, we perform a direct dark matter search using data from LIGO's first (O1) data run, as opposed to an indirect search for dark matter via its production of gravitational waves. We demonstrate an achieved sensitivity on squared coupling as $\sim4\times10^{-45}$, in a $U(1)_\text{B}$ dark photon dark matter mass band around $m_{\rm A}\sim 4 \times 10^{-13}$ eV. Substantially improved search sensitivity is expected during the coming years of continued data taking by LIGO and other gravitational wave detectors in a growing global network. 
}
\end{abstract}

\maketitle

\noindent{\bf \Large Introduction}

Although there is little doubt that dark matter (DM) exists in our Universe, its nature remains mysterious, including its component mass(es).
It may be an ultralight elementary particle, such as fuzzy DM with mass $\sim 10^{-22}$ eV \cite{Hu:2000ke,Marsh:2013ywa,Bozek:2014uqa,Hui:2016ltb}, or it may arise from stellar-mass objects, such as primordial black holes \cite{Carr:2009jm}.

One promising DM candidate in the ultralight mass regime is the dark photon (DP), which is the gauge boson of a $U(1)$ gauge group.
The DP can acquire its mass through the Higgs or Stueckelberg mechanism. As ultralight DM, the DP must be produced non-thermally, {\it e.g.}, production from the misalignment mechanism \cite{Nelson:2011sf,Arias:2012az,Graham:2015rva}, parametric resonance production or tachyonic instability of a scalar field \cite{Co:2018lka,Agrawal:2018vin,Bastero-Gil:2018uel,Dror:2018pdh}, or from the decay of a cosmic string network\cite{Long:2019lwl}.

It was recently proposed in \cite{Graham:2015ifn,Pierce:2018xmy} that a gravitational wave (GW) detector may be sensitive to dark photon dark matter (DPDM). The Advanced Laser Interferometer Gravitational-Wave Observatory (LIGO) consists of two 4-km dual-recycled Michelson Fabry-Perot interferometers in Livingston Louisiana (L1) and Hanford,
Washington (H1). From the first two observing runs (coincident with the Virgo detector for several weeks of the O2 run),
detections of ten binary black hole mergers and one binary neutron star merger have been reported~\cite{LIGOScientific:2018mvr}.
These measurements require a differential strain measurement sensitivity better than 10$^{-21}$ for broadband transients with
central frequencies of $O$(100 Hz), based on detecting minute changes in distance between the mirror pairs forming the
Fabry-Perot interferometer arms. 

Relevant to this search, the mirror separations can also change in response
to a gradient in a DPDM field due to non-zero photon velocity.
More explicitly, we consider a DP with mass $m_{\rm A}$ between $10^{-13}\sim10^{-11}$ eV.
The DPDM is an oscillating background field, for which the rest-frame oscillation
frequency satisfies: $f_0 \approx \left[{m_{\rm A}\over10^{-12}\>{\rm ev}}\right]$(241 Hz).
We assume the DP is the gauge boson of the gauged $U(1)_\text{B}$ group so that any object, including a LIGO mirror, that carries
baryon number will feel its oscillatory force, similar to that experienced by a macroscopic, electrically charged
object in an oscillating electric field. 

Using LIGO to look for DPDM bridges GW science and particle physics. In this paper, we present a $U(1)_\text{B}$ DPDM search using
data from Advanced LIGO's first observing run, O1. We confirm that the data from LIGO's first observation run yields results already more sensitive than limits from prior experiments in a narrow DPDM mass range. The sensitivity will be improved significantly with more LIGO data, as well as with the growth of the global network of GW detectors. Meanwhile, the same search strategy can be directly applied to search for many other ultralight DM scenarios, with excellent sensitivities achievable.

\vspace{0.5cm}

\noindent{\bf \Large Results}

%%%%%%%%%%%%%%%%%%%%%%%%%%%%%%%%%%%%%%%%%%%%%%%%%%%%%%%%%%%%%%%%%%%%
\begin{figure}[t]
\includegraphics[width=0.80\columnwidth, height=0.8\columnwidth]{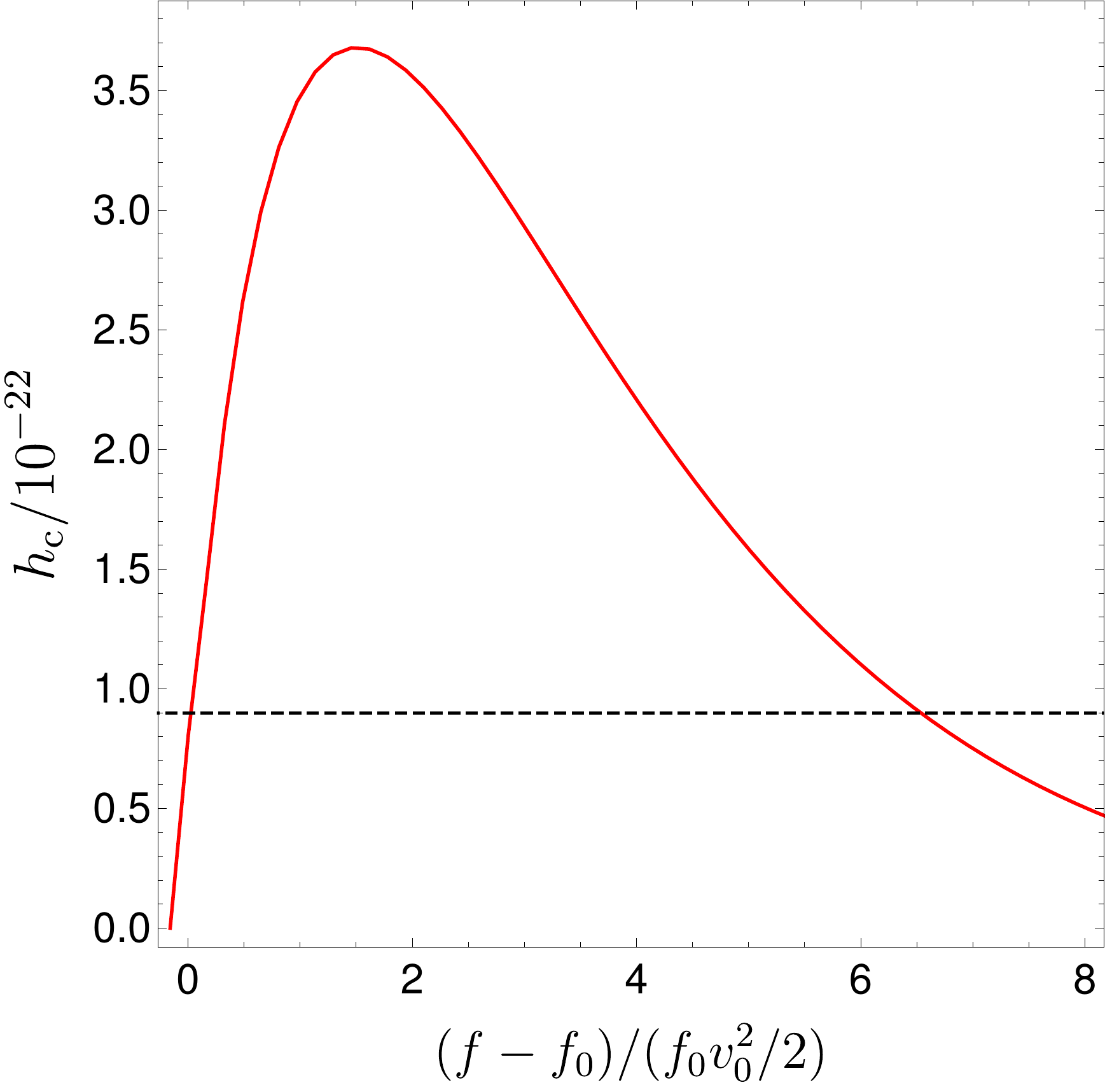}
  \caption{An example of dark photon dark matter signal power spectrum and corresponding detector sensitivity. The dark photon dark matter (DPDM) signal power spectrum is shown in terms of characteristic strains $h_{\rm c}$ (red), with $U(1)_{\rm B}$ coupling parameter $\epsilon^2 = 10^{-41}$, DPDM oscillation frequency $f_0=500$ Hz and typical velocity of DPDM $v_0=10^{-3}$ of the speed of light. The Advanced LIGO design sensitivity in a small frequency window is approximately flat, which is shown as the black dashed line.
\label{spectrum}
}
\end{figure}
%%%%%%%%%%%%%%%%%%%%%%%%%%%%%%%%%%%%%%%%%%%%%%%%%%%%%%%%%%%%%%%%%%%%

\noindent{\bf Estimating DPDM induced effects.}
Through virialization, DPDM particles in our galaxy have a typical velocity around $v_0 \equiv 10^{-3}$ of the speed of light, and thus they are highly non-relativistic. 
The total energy of a DM particle is then the sum of its mass energy and kinetic energy, i.e. $m_{\rm A}(1+v_0^2/2)$. Here and in the following, we use natural units, {\it i.e.} $c=\hslash=1$. Therefore the oscillation frequency of the DP field is approximately a constant, $\omega\simeq m_{\rm A}$, with $O(10^{-6})$ deviations.

Therefore within a small period of time and spatial separation, the DP field can be treated approximately as a planewave, {\it i.e.},
\begin{linenomath*}
\begin{eqnarray}\label{planewave}
A_\mu \simeq A_{\mu,0} \cos[m_{\rm A} t - \bf k\cdot\bf x+\theta].
\end{eqnarray}
\end{linenomath*}
Here $A_{\mu,0}$ is the amplitude of the DP field and $\theta$ is a random phase. The DP field strength can be simply written as $F_{\mu\nu}=\partial_\mu A_\nu-\partial_\nu A_\mu$. We choose the Lorenz gauge, $\partial^\mu A_\mu=0$, in what follows. In the non-relativistic limit, the dark electric field is much stronger than the dark magnetic field, and $A_t$ is negligible relative to $\bf{A}$. The magnitude of the DP field can be determined by the DM energy density, {\it i.e.}, 
$|{\bf A}_0| \simeq \sqrt{2\rho_{\text{DM}}}/m_{\rm A}$.

In Eq.~\ref{planewave}, we neglect the kinetic energy contribution to the oscillation frequency. We also set the polarization and propagation vectors, {\it i.e.}, ${\bf A}_0$ and $\bf k$, to be constant vectors.  This approximation is valid only when the observation is taken within the coherence region, {\it i.e.} $t_{\text{obs}}< t_{\text{coh}}\simeq \frac{4\pi}{m_{\rm A} v_{\text{vir}}^2}$ and $l_{\text{obs}}< l_{\text{coh}}\simeq \frac{2\pi}{m_{\rm A} v_{\text{vir}}}$. For example, if the DP field oscillates at 100 Hz, the coherence time is only $10^4$ s, much shorter than the total observation time. In order to model the DPDM field for a time much longer than the coherence time, we simulate the DPDM field by linearly adding up many planewaves propagating in randomly sampled directions. More details are given in the ``Methods'' section below.

From the DPDM background field ${\bf A}(t,{\bf x}_i)$, one can derive the acceleration induced by the DPDM on each test object, labeled by index $i$. This can be written:
\begin{linenomath*}
\begin{eqnarray}\label{eq:acc}
  {\bf a}_i(t, {\bf x}_i)\simeq\epsilon e \frac{q_{{\rm D},i}}{M_i}
  \partial_t {\bf A}(t,{\bf x}_i).
\end{eqnarray}
\end{linenomath*}
Here we use the approximation ${\bf E} \simeq \partial_t {\bf A}(t, {\bf x}_i)$ for the dark electric field. $q_{{\rm D},i}/M_i$ is the charge-mass-ratio of the test object in LIGO. Treating a DP as the gauge boson of $U(1)_\text{B}$, and given that the LIGO mirrors (test masses) are primarily silica, $q_{{\rm D},i}/M_i = 1/{\rm GeV}$. We note that results from \cite{Dror:2017ehi} impose very strong constraints on the gauged $U(1)_\text{B}$ scenario, due to gauge anomaly. However the results derived in these papers rely on an assumption of how to embed the model into a complete theory at high energy in order to cancel $U(1)_\text{B}$ anomalies. Extending the electroweak symmetry breaking sector or other anomaly cancellation mechanism can avoid such severe constraints. If one takes the model independent constraint on an anomalous gauge symmetry, new particles need to be added at energy scale $O(\frac{4\pi m_{\rm A}}{\epsilon e})$, which gives $O(\rm{TeV})$ for the parameter space we are interested in. Since the required new particles carry only electroweak charges, they are safe from various collider searches. We label the DP-baryon coupling as $\epsilon e$ where $e$ is the $U(1)_{\text{EM}}$ coupling constant. We emphasize that we choose DP to be a $U(1)_\text{B}$ gauge boson as a benchmark model. The same analysis strategy presented in this study can be directly applied to many other scenarios, such as a $U(1)_{\text{B}-\text{L}}$ gauge boson or scalar field which couples through Yukawa interactions. More details on various models, as well as subtleties when observation time is longer than coherence time, will be described in 
the future work.

\vspace{0.3cm}

\noindent{\bf Signal-to-Noise Ratio Estimation.}
We approximate the DPDM field as a planewave within a coherence region. For a DP field oscillating at frequency $O(100)$ Hz, the coherence length is $O(10^9\>{\rm m})$, much larger than the separation between the two LIGO GW detectors at Hanford and Livingston. Thus these two GW detectors experience a nearly identical DPDM field, inducing strongly correlated responses. Exploiting the correlation dramatically reduces the background in the analysis.

The DPDM signal is exceedingly narrowband, making Fourier analysis natural. We first compute discrete Fourier transforms (DFT) from
the time-domain data. The total observation time is broken into smaller, contiguous segments, each of duration $T_{\text{DFT}}$, with a total
observing time $T_{\text{obs}} = N_{\text{DFT}} T_{\text{DFT}}$. Denote the value of the complex DFT coefficient for two interferometers 1 and 2, DFT $i$ and frequency bin $j$ to be $z_{1(2),ij}$. The one-sided power spectral densities (PSD) for two interferometers are related to the raw powers as ${\rm PSD}_{1(2), ij} = 2P_{1(2), ij}/T_{\text{DFT}}$. $P_{1(2), ij}$ are taken to be the expectation values for $|z_{1(2),ij}|^2$, as estimated from neighboring, non-signal frequency bins, assuming locally flat noise (using a 50-bin running median estimate).

To an excellent approximation, the noise in the two LIGO interferometers is statistically independent, with the exception of particular
very narrow bands with electronic line disturbances~\cite{Covas:2018oik}, which are excluded from the analysis.
Detailed descriptions of broadband LIGO noise contributions may be found in~\cite{TheLIGOScientific:2016zmo}, including
discussion of potential environmental contaminations that could be correlated between the two
LIGO detectors, but none of which would mimic a DPDM signal.
The normalized signal strength using cross correlation of all simultaneous DFTs in the observation time can be written as 
\begin{linenomath*}
\begin{eqnarray}\label{signal}
S_j=\frac{1}{N_{\text{DFT}}}\sum_{i=1}^{N_{\text{DFT}}}\frac{z_{1,ij} z_{2,ij}^*}{P_{1,ij} P_{2,ij}}.
\end{eqnarray}
\end{linenomath*}
In the absence of a signal, the expectation value is zero and the variance of the real and imaginary parts is
\begin{linenomath*}
\begin{eqnarray}\label{noise}
\sigma^2_j=\frac{1}{N_{\text{DFT}}}\bigg\langle\frac{1}{2P_{1,j} P_{2,j}}\bigg\rangle_{N_{\text{DFT}}} ,
\end{eqnarray}
\end{linenomath*}
where $\langle\ \rangle_{N_{\text{DFT}}}$ denotes an average over the $N_{\text{DFT}}$ DFTs, which may have slowly varying non-stationarity.
The signal-to-noise ratio (SNR) can be defined by
\begin{linenomath*}
\begin{eqnarray}\label{SNR}
{\rm SNR}_j\equiv\frac{S_j}{\sigma_j}.
\end{eqnarray}
\end{linenomath*}
Taking into the account the small separation between the interferometers relative to the DP coherence length and their relative orientation (approximate 90-deg rotation of one interferometer's arms projected onto the other interferometer's plane), we expect the SNR$_j$ for a strong DPDM field to be primarily real and negative.

\vspace{0.3cm}

\noindent{\bf Efficiency factor.} In order to use the observed real(SNR) values to set limits on DPDM coupling as a function of frequency, one must correct for the signal power lost from binning. The suggested nominal binning proposed in \cite{Pierce:2018xmy} is $\Delta f/f = 10^{-6}$, based on a Maxwell velocity distribution. The binsize in frequency space is set by $T_{\text{DFT}}$, {\it i.e.}, $\Delta f = 1/T_{\text{DFT}}$, which is optimal at only $f_{\text{opt}}\simeq 10^6/T_{\text{DFT}}$. For a frequency higher than $f_{\text{opt}}$, the relative frequency binning is finer, implying loss of signal power in single-bin measurements. At frequencies lower than $f_{\text{opt}}$, the relative frequency binning is coarser, implying full capture of the signal power, but at the cost of unnecessarily increased noise. We note that it is possible the DM velocity distribution deviates from Maxwell distribution by an $O(1)$ factor, e.g. \cite{Lentz:2017aay,Necib:2018iwb}. However, the impact is small, as the single-bin search used here depends on the integrated power within a frequency bin and not so much on its shape.

In Fig.~\ref{spectrum}, we show the DPDM signal power spectrum as a function of frequency, where $f_0=m_{\rm A}/2\pi$. We choose to normalize the x-axis by the intrinsic signal width, determined by the typical kinetic energy of DM particles. In this calculation, we include the Earth rotation effect. Without it, the signal PSD is proportional to $v f(v)$ where $f(v)$ is the Maxwell distribution. The Earth's rotation broadens our signal by  $ \Delta f \approx 2 f_\text{E}$. Different choices of $f_0$ result in slightly different deformations after including the rotation, but the changes are negligible in the frequency regime of interest. An analytical understanding of the PSD will be presented in the future work.

\begin{figure}[t]
\includegraphics[width=0.99\columnwidth]{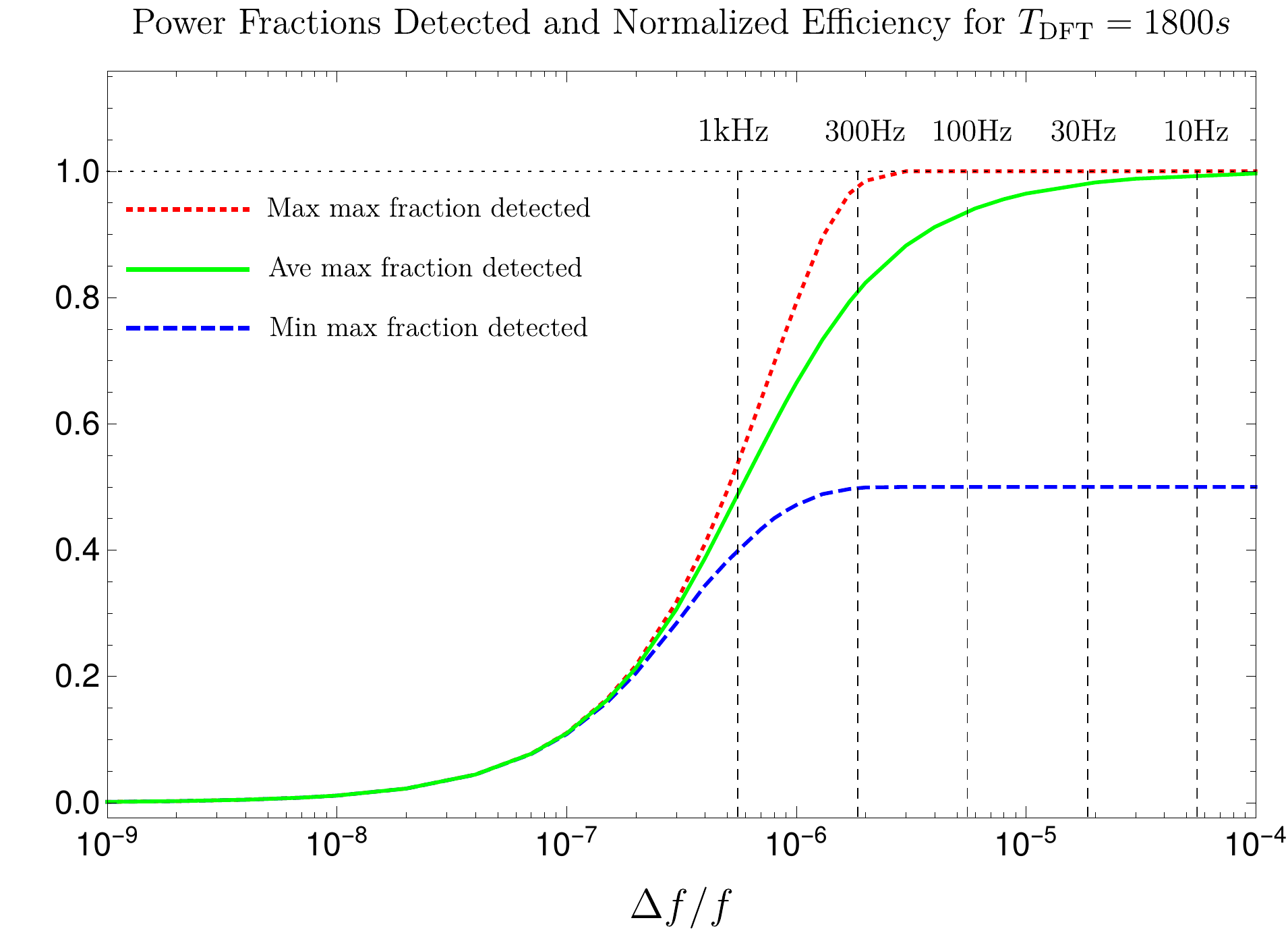}
\caption{ Signal power single-bin detection efficiency as a function of relative frequency resolution for a fixed
  coherence time of 1800 s. The upper (red) curve is for an optimal bin boundary choice
  ({\it a priori} unknown) for a given signal. The lower (blue) curve shows the worst-case
  efficiency for the least optimal boundary choice. The middle (green) curve shows an average
  over randomly chosen boundary choices.
}
\label{efficiency}
\end{figure}

The power spectrum from numerical simulation is used to determine empirically the fractions of power falling into a single fixed $\Delta f/f$ bin, where bin boundaries are systematically varied over the allowed range. Fig. \ref{efficiency} shows the resulting efficiencies (power fractions) for $T_{\text{DFT}}$ set to be 1800s. The red 
dotted curve shows the best case, for which the bin boundary is optimal. The blue dashed curve shows the worst case, which necessarily approaches $50\%$ for coarse binning (low frequency), while the green solid curve shows the average maximum efficiency over all bin boundary choices. A fit to the green solid curve is used for deriving upper limits on DPDM coupling. 

%%%%%%%%%%%%%%%%%%%%%%%%%%%%%%%%%%%%%%%%%%%%%%%%%%%%%%%
%%%%%%%%%%%%%%%%%%%%%%%%%%%%%%%%%%%%%%%%%%%%%%%%%%%%%%%

\vspace{0.3cm}

\noindent{\bf Data selection and analysis.}
The strain data used in this analysis was downloaded from the Gravitational Wave Open Science Center (GWOSC) web site~\cite{Vallisneri:2014vxa} and transformed to create
1786 1800-second coincident \text{DFT}s from the L1 and H1 interferometers. The GWOSC data sets exclude short periods 
during which overall data quality is poor. The choice of coherence time in this first DPDM search is somewhat arbitrary,
but allowed convenient comparison of spectral line artifacts observed with those reported from 1800-s \text{DFT}s in
LIGO continuous gravitational wave searches, for which 1800-s is a common \text{DFT} duration chosen. A shorter coherence 
time would be more optimal at frequencies above $\sim 500$ Hz for this single-bin detection analysis. In principle, a longer time would be more optimal
for lower frequencies, but in practice, sporadic interruptions of interferometer operations during data taking
lead to significant livetime loss for DFTs requiring very long contiguous periods of coincident Hanford-Livingston operations.

%%%%%%%%%%%%%%%%%%%%%%%%%%%%%%%%%%%%%%%%%%%%%%%%%%%%%%%
\begin{figure*}
\includegraphics[width=0.95\textwidth]{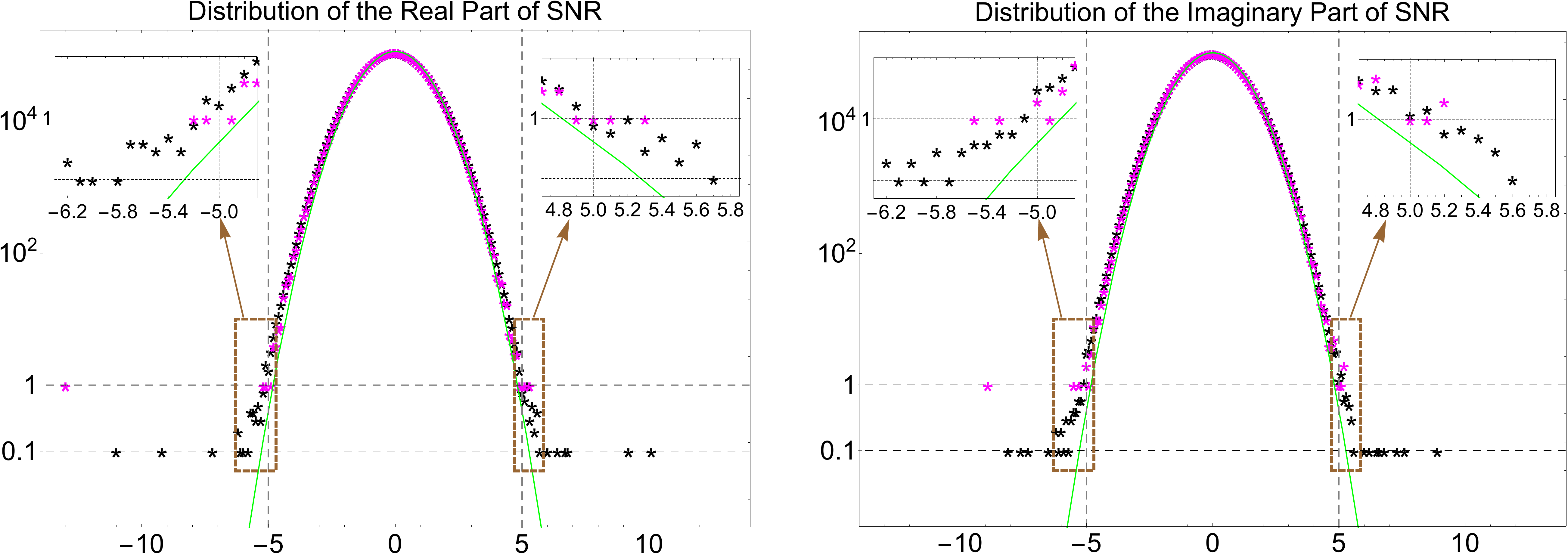}
\caption{Distributions of the real and imaginary parts of the signal-to-noise ratio. 
  The signal-to-noise ratio (SNR) for the signal bins (``zero lag'') are labeled in magenta and the lagged (control) bins in black, along with the ideal Gaussian expectation in green.}
\label{SNRdist}
\end{figure*}
%%%%%%%%%%%%%%%%%%%%%%%%%%%%%%%%%%%%%%%%%%%%%%%%%%%%%%%

The search for detections and the setting of upper limits in the absence of detection is based on ``loud'' values of
the detection statistic~(Eq.~\ref{SNR}). Specifically, we look for large negative real values of the SNR. Since
we search over $\sim4$ million DFT bins in the band 10-2000 Hz, we must correct for a large statistical trials factor in assessing
what SNR value is deemed ``significant.'' We choose a nominal signal candidate selection of SNR $< -5.8$, corresponding to a $\sim1$\%\
false alarm probability, assuming Gaussian noise. In practice, the noise in some frequency bands is not truly Gaussian,
leading to excess counts at large SNR. To assess the severity of this effect, we also define and examine
control bands (``frequency lags'') in which a DFT frequency bin in one interferometer is compared to a set of
offset bins from the other interferometer such that a true DPDM signal would not contribute to a non-zero cross correlation,
but for which single-interferometer artifacts or broadband correlated artifacts lead to non-zero correlation.
This frequency lag method is analogous to the time lag method used in transient gravitational wave analysis. Specifically, we choose 10 lags
of ($-50$, $-40$, ..., $-10$, $+10$, ..., $+50$) frequency bin offsets to assess the non-Gaussian background from these instrumental artifacts.
To avoid contamination of both signal and control bands from known artifacts, we exclude from the analysis any band within $\sim0.056$ Hz of
a narrow disturbance listed in~\cite{Covas:2018oik}, where the extra veto margin is to reduce susceptibility to spectral leakage from strong lines. We also exclude the band 331.3-331.9 Hz, for which extremely loud narrow calibration
excitations in the two interferometers lead to significant overlapping spectral leakage and hence non-random correlation.

Fig.~\ref{SNRdist} shows the distributions of the real and imaginary parts of the SNR~(Eq.~\ref{SNR}) for both the
signal bins (``zero lag'') in magenta and the lagged bins in black. The distributions follow quite closely the ideal
Gaussian curve shown, except for a slight excess visible in the tails beyond $|{\rm SNR}|>5$ (note there are $\sim 10$ times
as many lagged bins as signal bins in the graphs). The only signal bins with $|{\rm SNR}|>5.8$ arise from known
continuous wave ``hardware injections'' used in detector response validation, for which the complex SNR can have an arbitrary
phase in the cross correlation that depends on the simulated source frequency and direction. An investigation was carried out of all other SNR outliers (10) with real or imaginary values having magnitudes greater than 5. In all but three cases, lagged bins in neighboring bins within 0.2 Hz of the signal bin showed elevated noise, defined by an SNR magnitude greater than 4, suggesting non-Gaussian contamination.
The Gaussian-noise expectation for this range [5.0-5.8] of subthreshold outlier magnitude (real or imaginary) is 4.1 events, consistent with observation
in clean bands.

Since no significant candidates were found, upper limits were set. In future searches, should significant candidates appear, it will
be critical to assess their consistency with instrumental artifacts. A simple approach is to increase the number of control bins
examined per candidate, to assess better potential non-Gaussian single-interferometer contamination and broadband correlated artifacts.
A greater concern would be a highly narrowband correlated disturbance, such as from identical electronic instruments at each
observatory creating a sharp spectral line through electrical current draws in power supplies affecting interferometer controls.
Detailed investigation using auxiliary instrumental and environmental channels would be warranted, to exclude such interference.

\noindent{\bf  New constraints from LIGO O1 data. } 
\begin{figure*}
\includegraphics[width=0.7\textwidth]{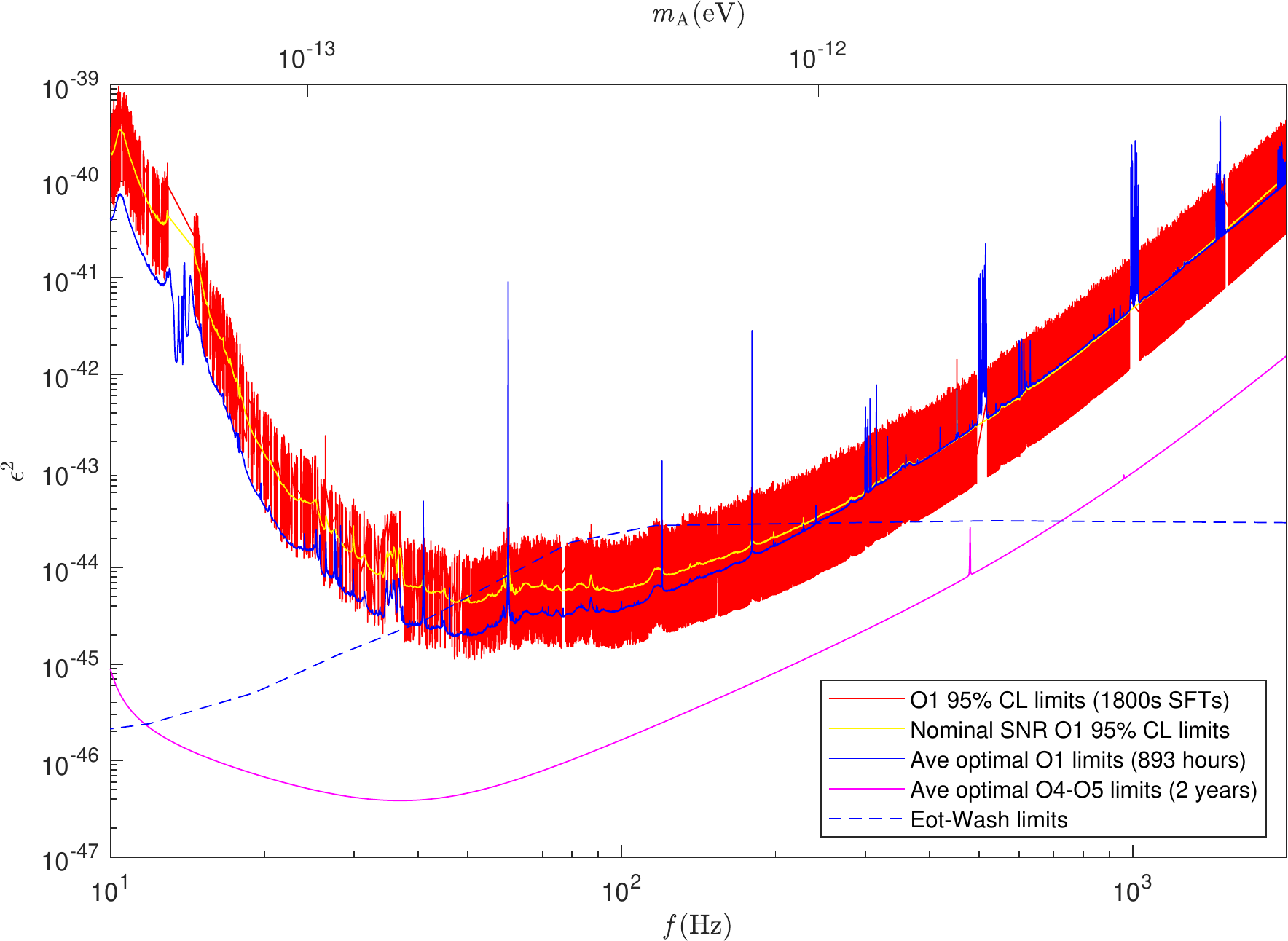}
\caption{Derived $95\%$ confidence level upper limits on the coupling parameter $\epsilon^2$ for dark photon-baryon coupling. The broad red band shows the actual upper limits with $1/1800$ Hz binning. The yellow curve shows the expected upper limit for an average measured real(SNR) = 0. The dark blue curve shows the ``optimal'' upper limit expected when the discrete Fourier transform (DFT) binning adjusts with frequency to maintain $\Delta f/f = 10^{-6}$ for the same 893-hour observation time. The magenta curve shows the ``optimal'' upper limit expected for a 2-year, $100\%$-livetime run at Advanced LIGO design sensitivity (``O4-O5"). The dashed curve shows upper limits derived from the E$\ddot{\textrm{o}}$t-Wash group \cite{Su:1994gu,Schlamminger:2007ht}. This is a fifth-force experiment, whose constraint does not rely on dark photon (DP) being dark matter (DM). The large spikes of red and blue curves, overlaid on top of each other, are induced by known sources of noise, such as vibrations of mirror suspension fibers. }\label{result}
\end{figure*}

Our main results are presented in Fig.~\ref{result}.  We show the derived $95\%$ confidence level (CL) upper limits on the parameter $\epsilon^2$ for DP-baryon coupling, as a function of DPDM oscillating frequency. The broad red band shows the range of upper limits obtained with $1/1800$ Hz binning, using the measured real part of the SNR detection statistic defined below and the Feldman-Cousins (FC) formalism~\cite{Feldman:1997qc} and after applying an efficiency correction discussed below. The yellow curve shows the expected upper limit for an average measured real(SNR) = 0, applying the same FC formalism and efficiency correction. The dark blue curve shows a more optimal upper limit expected when the DFT binning adjusts with frequency to maintain $\Delta f/f = 10^{-6}$ for the same 893-hour observation time, for the same efficiency correction and for an averaged detector sensitivity equal to that in the analysis. The yellow and dark blue curves agree well with each other at around 500 Hz, where $1/1800$ Hz is the optimal choice of the binsize. The mean achieved upper limit is generally worse than the optimal sensitivity because, with fixed binsize at $1/1800$ Hz, excess noise is included at low frequency and some signal power is lost at high frequency. The dashed curve shows upper limits derived from the E$\ddot{\textrm{o}}$t-Wash group based on Equivalence Principle tests using a torsion balance~\cite{Su:1994gu,Schlamminger:2007ht}. Given the LIGO O1 data, under the assumption that DP constitutes all DM, we have already improved upon existing limits in a mass window around $m_{\rm A}\sim 4\times10^{-13}$ eV.

Future searches in more sensitive data will probe deeper into an unexplored $\epsilon^2$--$m_{\rm A}$ parameter space. Assuming no discovery and a negligible true GW stochastic background,
the magenta curve shows the ``optimal'' upper limit expected for a 2-year, $100\%$-livetime run at Advanced LIGO design sensitivity (``O4-O5"). This limit looks smoother, as it uses a design sensitivity curve that shows only fundamental noise sources, while the blue curve includes additional, non-fundamental noise artifacts that have not yet been mitigated in LIGO detector commissioning, such as power mains contamination at 60 Hz and harmonics and environmental vibrations. The simulations discussed below uncovered an error of a factor of 4 in the $\epsilon^2$--$m_{\rm A}$ sensitivity plot in \cite{Pierce:2018xmy}. This error has been corrected in the current study. As GW detectors become more sensitive in the future, one expects a stochastic GW background from compact binary coalescence mergers to emerge eventually, with an
 integrated broadband stochastic signal perhaps detectable as early as the O4-O5 run \cite{LIGOScientific:2019vic}.
Nonetheless, the stochastic GW strain power from mergers in a single DPDM search bin will remain negligible for years to come.

The inclusion of a global network of detectors, such as Virgo, KAGRA and LIGO-India, improves the DPDM search sensitivity, in principle. The degree of 
improvement depends, however, upon the relative alignments among these detectors as well as their sensitivities.
The Virgo detector is currently less sensitive than the two LIGO detectors. In addition, its orientation is not well aligned with those of the LIGO detectors. Future third-generation detectors, such as Einstein Telescope and Cosmic Explorer, will have much lower noise, permitting still more sensitive
DPDM searches.

\vspace{0.5cm}

\noindent{\bf \Large Discussion}

In this paper, we present a direct DM search using gravitational wave detector strain data. We choose the $U(1)_\text{B}$ DP as our benchmark scenario; our early results already improve upon prior searches in a narrow DP mass range, and future searches will probe deeper in DPDM coupling strength and wider in mass range.
This first analysis uses a non-templated, single-Fourier-bin cross-correlation detection statistic.
Refinements to be examined for analysis of future data sets include multiple DFT coherence times, tuned according to search band,
templated filtering over multiple Fourier bins and exploitation of extremely narrow features expected in the DPDM spectrum, resolvable by GW detectors for loud enough SNR.

With more data to be collected by LIGO and other gravitational wave detectors in the coming years, as well as with improved search strategies,
we expect DPDM searches to probe steadily deeper in DPDM parameter space. The same strategy can be implemented directly in searches for many other ultralight DM scenarios.  This novel use of data from a gravitational wave detector demonstrates the versatility of these remarkable instruments for directly probing exotic physics.

\vspace{0.5cm}

\noindent{\bf \Large Methods}

\noindent {\bf Simulating the DPDM background.} 
The DPDM background is a superposition of many DP wave functions, similar to the axion DM background as studied in \cite{Foster:2017hbq}. In the galaxy frame, each 
DP has a random polarization direction isotropically distributed. The magnitude of $\bf A$ is taken to be a fixed number for each DP particle
with normalization discussed below. As for the polarization vector, the velocity direction also follows an isotropic
distribution. The magnitude of the velocity is obtained from the Maxwell distribution
\begin{linenomath*}
\begin{equation}\label{velodist}
f(v)\sim v^2e^{-v^2/v^2_0},
\end{equation}
\end{linenomath*}
where $v_0$ is taken to be $0.77\times10^{-3} c$ \cite{Smith:2006ym}. In the non-relativistic limit, the polarization vector and the velocity vector are independent of each other.

For the $i$-th DP particle, the wave function can be written as 
\begin{linenomath*}
\begin{eqnarray}\label{vector}
  {\bf A}_i(t, \mathbf{x}) \equiv{\bf A}_{i,0} \sin(\omega_i t - {\bf k}_i \cdot {\bf x}+\phi_{i}), 
\end{eqnarray}
\end{linenomath*}
where $\omega_i = \sqrt{{\bf k}_i^2 + m_{\rm A}^2} \equiv 2 \pi f_i$ and ${\bf k}_i = m_{\rm A} {\bf v}_i$. The DPDM background can be generated by superposing many, $N$, of these 
wave functions 
\begin{linenomath*}
\begin{eqnarray}\label{Atot}
  {\bf A}_{\text{total}}(t, \mathbf{x}) =\sum_{i=1}^N {\bf A}_{i, 0} \sin(\omega_i t - {\bf k}_i \cdot {\bf x}+\phi_{i}) .
\end{eqnarray}
\end{linenomath*}
Here the phase of the wave function for each particle, $\phi_i$, is randomly chosen from a uniform distribution from 0 to $2\pi$. 

To simulate DPDM background, we consider $N=10^3$ DPDM particles. We note that having $N=10^3$ suffices to reveal essential features of the DPDM background, such as coherence length and coherence time. Further, this simulation provides a signal PSD which agrees well with analytic results ($N\to \infty$ limit). Thus we believe that the result from this simulation is reliable. 

Finally, the normalization of ${\bf A}_{i,0}$ is determined by the local DM energy density. In the non-relativistic limit, the energy density of DM can be calculated as
\begin{linenomath*}
\begin{equation}
  \frac{1}{V}\frac{1}{T}\int_{V} d^3x\int_{0}^{T} dt\ m_{\rm A}^2{\bf A}_{\text{total}}^2=\rho_{\text{DM}}\simeq 0.4\ \rm{GeV/cm}^3.  
\end{equation}
\end{linenomath*}
In order to average out the fluctuations in numerical simulation, the temporal and spatial integrals are taken to be much longer than
the coherence time and length, {\it i.e.}, $T\gg T_{\text{coh}}$ and $V\gg l_{\text{coh}}^3$. Since the DPDM is obtained from a superposition of $N$ DP particles in an uncorrelated manner, the total amplitude increases as $\sqrt{N}$. For a fixed DM energy density $\rho_{\text{DM}}$, one expects $|{\bf A}_{i,0}|\simeq \sqrt{2\rho_{\text{DM}}}/(m_{\rm A}\sqrt{N})$, consistent with our numerical results.

\noindent{\bf Interface to LIGO simulations.} We use the LIGO Scientific Collaboration Algorithm Library Suite (LALSuite) \cite{LAL} for mimicking of GW detector response of DPDM and for superposing random Gaussian noise. This suite of programs has been developed over two decades for simulating GW signals, detector response and for carrying out GW analysis, including source parameter estimation. 

Below, we give a brief overview of the relevant LALSuite GW response model and explain what is modified to simulate DPDM induced effects.
When the GW wavelength is much longer than the detector's characteristic size, {\it i.e.}, $\lambda\gg L$, one can use the equation of the geodesic deviation in the proper detector frame to calculate the GW induced effect, 
\begin{linenomath*}
\begin{equation}
\ddot{\xi}^i=\frac{1}{2}\ddot{h}_{ij}^{\rm TT}\xi^j,
\end{equation}
\end{linenomath*}
where $\xi$ is the coordinate of a test object in the proper detector frame. At leading order, the relative change of the arm length is
\begin{linenomath*}
\begin{equation}
  R\equiv \frac{\Delta L_\text{x} - \Delta L_\text{y}}{L} = h^+ F_+ + h^\times F_\times  ,
\end{equation}
\end{linenomath*}
where $F_+$ and $F_\times$ are antenna pattern functions, which can be written as
\begin{linenomath*}
\begin{eqnarray}
  F_+&=&\sum_{i,j} D_{ij}({\bf e_+})_{ij}=\frac{1}{2}\left[({\bf e}_+)_{xx}-({\bf e}_+)_{yy}\right],\nonumber\\
  F_\times&=&\sum_{i,j} D_{ij}({\bf e}_\times)_{ij}=\frac{1}{2}\left[({\bf e}_\times)_{xx}-({\bf e}_\times)_{yy}\right] ,
\end{eqnarray}
\end{linenomath*}
with polarization tensors
\begin{linenomath*}
\begin{eqnarray}
({\bf e}_+)_{ij}&=&({\bf X}\otimes{\bf X}-{\bf Y}\otimes{\bf Y})_{ij},\nonumber\\
({\bf e}_\times)_{ij}&=&({\bf X}\otimes{\bf Y}+{\bf Y}\otimes{\bf X})_{ij},
\end{eqnarray}
\end{linenomath*}
and detector tensors
\begin{linenomath*}
\begin{equation}
  D_{ij}=\frac{1}{2}({\bf n}^\text{x}\otimes{\bf n}^\text{x}-{\bf n}^\text{y}\otimes{\bf n}^\text{y})_{ij},
\end{equation}
\end{linenomath*}
where vectors ${\bf X}$ and ${\bf Y}$ are the axes of the wave frame, and ${\bf n}^{\text{x}}$ and ${\bf n}^\text{y}$ are unit vectors along the $x$ and $y$ arms of LIGO respectively.

In order to concretely estimate LIGO's sensitivity to a DPDM signal, we calculate the DPDM induced relative change of the arm length as a function of time, {\it i.e.}, $R(t)$. Then we inject this as the signal into LALSuite. The background is further added as a Gaussian white noise. As a benchmark, the 
DP oscillation frequency is set to be $100/\sqrt{2}$ Hz and $\epsilon^2$ to be $5\times 10^{-44}$. For the simulation, we take 
$T_{\text{DFT}}=1800$ s, $T_{\text{obs}}=200$ hours and $\sqrt{\rm PSD}=10^{-23}/\sqrt{\rm{Hz}}$. The signal appears as a negative real number, {\it i.e.}, SNR $\simeq -8$. The sensitivity to $\epsilon^2$ scales as 
\begin{linenomath*}
\begin{equation}\label{scale}
  \frac{\epsilon'^2}{\epsilon^2}=\frac{\rm{SNR}'}{\rm{SNR}}\frac{T_{\text{coh}}}{T'_{\text{coh}}}\sqrt{\frac{N_{\text{DFT}}}{N'_{\text{DFT}}}}.
\end{equation}
\end{linenomath*}
With this scaling, our simulations are consistent with upper limits shown in Fig. \ref{result} based on the search of O1 data.

\vspace{0.5cm}

\noindent{\bf \Large Data availability}

The upper limits in Fig.~\ref{result} can be found at https://doi.org/10.5281/zenodo.3525343 and
the LIGO O1 data that support the findings of this study are available from the GWOSC website https://www.gw-openscience.org~\cite{Vallisneri:2014vxa}.

\vspace{0.5cm}

\noindent{\bf \Large Code availability}

The source code for the analysis is available from the corresponding author upon request.

\vspace{0.5cm}

\noindent{\bf \Large Acknowledgements}

\noindent
The authors thank the members of the LIGO Scientific Collaboration and Virgo Collaboration for useful discussions,
and thank Ed Daw for helpful comments on this manuscript.
This research has made use of data obtained from the Gravitational Wave Open Science Center (https://www.gw-openscience.org), a service of LIGO Laboratory, the LIGO Scientific Collaboration and the Virgo Collaboration. LIGO is funded by the U.S. National Science Foundation. Virgo is funded by the French Centre National de Recherche Scientifique (CNRS), the Italian Istituto Nazionale della Fisica Nucleare (INFN) and the Dutch Nikhef, with contributions by Polish and Hungarian institutes.
This work was partly funded by National Science Foundation grants NSF PHY 1505932 and NSF PHY 1806577. H.K.G. is supported in part by the U.S. Department of Energy grant DE-SC0009956. F.W.Y. thanks the host of the University of Utah while this work is being finished. Y.Z. is supported by the U.S. Department of Energy under Award Number DE-SC0009959.

\vspace{0.5cm}

\noindent{\bf \Large Author contributions}

\noindent
H.K.G., K.R., F.W.Y. and Y.Z. have contributed equally to the intellectual content of the paper and
the design of the analysis.

\vspace{0.3cm}

\noindent{\bf \Large Competing interests}

\noindent
The authors declare no competing interests.


\vspace{0.5cm}

\noindent{\bf \Large References}


\begin{thebibliography}{99}
\expandafter\ifx\csname url\endcsname\relax
  \def\url#1{\texttt{#1}}\fi
\expandafter\ifx\csname urlprefix\endcsname\relax\def\urlprefix{URL }\fi
\providecommand{\bibinfo}[2]{#2}
\providecommand{\eprint}[2][]{\url{#2}}

\bibitem{Hu:2000ke}
\bibinfo{author}{Hu, W.}, \bibinfo{author}{Barkana, R.} \&
  \bibinfo{author}{Gruzinov, A.}
\newblock \bibinfo{title}{{Cold and fuzzy dark matter}}.
\newblock \emph{\bibinfo{journal}{Phys. Rev. Lett.}}
  \textbf{\bibinfo{volume}{85}}, \bibinfo{pages}{1158--1161}
  (\bibinfo{year}{2000}).

\bibitem{Marsh:2013ywa}
\bibinfo{author}{Marsh, D. J.~E.} \& \bibinfo{author}{Silk, J.}
\newblock \bibinfo{title}{{A Model For Halo Formation With Axion Mixed Dark
  Matter}}.
\newblock \emph{\bibinfo{journal}{Mon. Not. Roy. Astron. Soc.}}
  \textbf{\bibinfo{volume}{437}}, \bibinfo{pages}{2652--2663}
  (\bibinfo{year}{2014}).

\bibitem{Bozek:2014uqa}
\bibinfo{author}{Bozek, B.}, \bibinfo{author}{Marsh, D. J.~E.},
  \bibinfo{author}{Silk, J.} \& \bibinfo{author}{Wyse, R. F.~G.}
\newblock \bibinfo{title}{{Galaxy UV-luminosity function and reionization
  constraints on axion dark matter}}.
\newblock \emph{\bibinfo{journal}{Mon. Not. Roy. Astron. Soc.}}
  \textbf{\bibinfo{volume}{450}}, \bibinfo{pages}{209--222}
  (\bibinfo{year}{2015}).

\bibitem{Hui:2016ltb}
\bibinfo{author}{Hui, L.}, \bibinfo{author}{Ostriker, J.~P.},
  \bibinfo{author}{Tremaine, S.} \& \bibinfo{author}{Witten, E.}
\newblock \bibinfo{title}{{Ultralight scalars as cosmological dark matter}}.
\newblock \emph{\bibinfo{journal}{Phys. Rev.}} \textbf{\bibinfo{volume}{D95}},
  \bibinfo{pages}{043541} (\bibinfo{year}{2017}).

\bibitem{Carr:2009jm}
\bibinfo{author}{Carr, B.~J.}, \bibinfo{author}{Kohri, K.},
  \bibinfo{author}{Sendouda, Y.} \& \bibinfo{author}{Yokoyama, J.}
\newblock \bibinfo{title}{{New cosmological constraints on primordial black
  holes}}.
\newblock \emph{\bibinfo{journal}{Phys. Rev.}} \textbf{\bibinfo{volume}{D81}},
  \bibinfo{pages}{104019} (\bibinfo{year}{2010}).

\bibitem{Nelson:2011sf}
\bibinfo{author}{Nelson, A.~E.} \& \bibinfo{author}{Scholtz, J.}
\newblock \bibinfo{title}{{Dark Light, Dark Matter and the Misalignment
  Mechanism}}.
\newblock \emph{\bibinfo{journal}{Phys. Rev.}} \textbf{\bibinfo{volume}{D84}},
  \bibinfo{pages}{103501} (\bibinfo{year}{2011}).

\bibitem{Arias:2012az}
\bibinfo{author}{Arias, P.} \emph{et~al.}
\newblock \bibinfo{title}{{WISPy Cold Dark Matter}}.
\newblock \emph{\bibinfo{journal}{JCAP}} \textbf{\bibinfo{volume}{1206}},
  \bibinfo{pages}{013} (\bibinfo{year}{2012}).

\bibitem{Graham:2015rva}
\bibinfo{author}{Graham, P.~W.}, \bibinfo{author}{Mardon, J.} \&
  \bibinfo{author}{Rajendran, S.}
\newblock \bibinfo{title}{{Vector Dark Matter from Inflationary Fluctuations}}.
\newblock \emph{\bibinfo{journal}{Phys. Rev.}} \textbf{\bibinfo{volume}{D93}},
  \bibinfo{pages}{103520} (\bibinfo{year}{2016}).

\bibitem{Co:2018lka}
\bibinfo{author}{Co, R.~T.}, \bibinfo{author}{Pierce, A.},
  \bibinfo{author}{Zhang, Z.} \& \bibinfo{author}{Zhao, Y.}
\newblock \bibinfo{title}{{Dark Photon Dark Matter Produced by Axion
  Oscillations}}.
\newblock \emph{\bibinfo{journal}{Phys. Rev.}} \textbf{\bibinfo{volume}{D99}},
  \bibinfo{pages}{075002} (\bibinfo{year}{2019}).

\bibitem{Agrawal:2018vin}
\bibinfo{author}{Agrawal, P.}, \bibinfo{author}{Kitajima, N.},
  \bibinfo{author}{Reece, M.}, \bibinfo{author}{Sekiguchi, T.} \&
  \bibinfo{author}{Takahashi, F.}
\newblock \bibinfo{title}{{Relic Abundance of Dark Photon Dark Matter.}}
  Preprint at https://arxiv.org/abs/1810.07188
  (\bibinfo{year}{2018}).

\bibitem{Bastero-Gil:2018uel}
\bibinfo{author}{Bastero-Gil, M.}, \bibinfo{author}{Santiago, J.},
  \bibinfo{author}{Ubaldi, L.} \& \bibinfo{author}{Vega-Morales, R.}
\newblock \bibinfo{title}{{Vector dark matter production at the end of
  inflation}}.
\newblock \emph{\bibinfo{journal}{JCAP}} \textbf{\bibinfo{volume}{1904}},
  \bibinfo{pages}{015} (\bibinfo{year}{2019}).

\bibitem{Dror:2018pdh}
\bibinfo{author}{Dror, J.~A.}, \bibinfo{author}{Harigaya, K.} \&
  \bibinfo{author}{Narayan, V.}
\newblock \bibinfo{title}{{Parametric Resonance Production of Ultralight Vector
  Dark Matter}}.
\newblock \emph{\bibinfo{journal}{Phys. Rev.}} \textbf{\bibinfo{volume}{D99}},
  \bibinfo{pages}{035036} (\bibinfo{year}{2019}).

\bibitem{Long:2019lwl}
\bibinfo{author}{Long, A.~J.} \& \bibinfo{author}{Wang, L.-T.}
\newblock \bibinfo{title}{{Dark Photon Dark Matter from a Network of Cosmic
  Strings}}.
\newblock \emph{\bibinfo{journal}{Phys. Rev.}} \textbf{\bibinfo{volume}{D99}},
  \bibinfo{pages}{063529} (\bibinfo{year}{2019}).

\bibitem{Graham:2015ifn}
\bibinfo{author}{Graham, P.~W.}, \bibinfo{author}{Kaplan, D.~E.},
  \bibinfo{author}{Mardon, J.}, \bibinfo{author}{Rajendran, S.} \&
  \bibinfo{author}{Terrano, W.~A.}
\newblock \bibinfo{title}{{Dark Matter Direct Detection with Accelerometers}}.
\newblock \emph{\bibinfo{journal}{Phys. Rev.}} \textbf{\bibinfo{volume}{D93}},
  \bibinfo{pages}{075029} (\bibinfo{year}{2016}).

\bibitem{Pierce:2018xmy}
\bibinfo{author}{Pierce, A.}, \bibinfo{author}{Riles, K.} \&
  \bibinfo{author}{Zhao, Y.}
\newblock \bibinfo{title}{{Searching for Dark Photon Dark Matter with
  Gravitational Wave Detectors}}.
\newblock \emph{\bibinfo{journal}{Phys. Rev. Lett.}}
  \textbf{\bibinfo{volume}{121}}, \bibinfo{pages}{061102}
  (\bibinfo{year}{2018}).

\bibitem{LIGOScientific:2018mvr}
\bibinfo{author}{Abbott, B.~P.} \emph{et~al.}
\newblock \bibinfo{title}{{GWTC-1: A Gravitational-Wave Transient Catalog of
  Compact Binary Mergers Observed by LIGO and Virgo during the First and Second
  Observing Runs}}.
\newblock \emph{\bibinfo{journal}{Phys. Rev.}} \textbf{\bibinfo{volume}{X9}},
  \bibinfo{pages}{031040} (\bibinfo{year}{2019}).

\bibitem{Dror:2017ehi}
\bibinfo{author}{Dror, J.~A.}, \bibinfo{author}{Lasenby, R.} \&
  \bibinfo{author}{Pospelov, M.}
\newblock \bibinfo{title}{{New constraints on light vectors coupled to
  anomalous currents}}.
\newblock \emph{\bibinfo{journal}{Phys. Rev. Lett.}}
  \textbf{\bibinfo{volume}{119}}, \bibinfo{pages}{141803}
  (\bibinfo{year}{2017}).

\bibitem{Covas:2018oik}
\bibinfo{author}{Covas, P.~B.} \emph{et~al.}
\newblock \bibinfo{title}{{Identification and mitigation of narrow spectral
  artifacts that degrade searches for persistent gravitational waves in the
  first two observing runs of Advanced LIGO}}.
\newblock \emph{\bibinfo{journal}{Phys. Rev.}} \textbf{\bibinfo{volume}{D97}},
  \bibinfo{pages}{082002} (\bibinfo{year}{2018}).

\bibitem{TheLIGOScientific:2016zmo}
\bibinfo{author}{Abbott, B.~P.} \emph{et~al.}
\newblock \bibinfo{title}{{Characterization of transient noise in Advanced LIGO
  relevant to gravitational wave signal GW150914}}.
\newblock \emph{\bibinfo{journal}{Class. Quant. Grav.}}
  \textbf{\bibinfo{volume}{33}}, \bibinfo{pages}{134001}
  (\bibinfo{year}{2016}).

\bibitem{Lentz:2017aay}
\bibinfo{author}{Lentz, E.~W.}, \bibinfo{author}{Quinn, T.~R.},
  \bibinfo{author}{Rosenberg, L.~J.} \& \bibinfo{author}{Tremmel, M.~J.}
\newblock \bibinfo{title}{{A New Signal Model for Axion Cavity Searches from
  N-Body Simulations}}.
\newblock \emph{\bibinfo{journal}{Astrophys. J.}}
  \textbf{\bibinfo{volume}{845}}, \bibinfo{pages}{121} (\bibinfo{year}{2017}).

\bibitem{Necib:2018iwb}
\bibinfo{author}{{Necib}, L.}, \bibinfo{author}{{Lisanti}, M.} \&
  \bibinfo{author}{{Belokurov}, V.}
\newblock \bibinfo{title}{{Inferred Evidence for Dark Matter Kinematic
  Substructure with SDSS-Gaia}}.
\newblock \emph{\bibinfo{journal}{\apj}} \textbf{\bibinfo{volume}{874}},
  \bibinfo{pages}{3} (\bibinfo{year}{2019}).

\bibitem{Vallisneri:2014vxa}
\bibinfo{author}{Vallisneri, M.}, \bibinfo{author}{Kanner, J.},
  \bibinfo{author}{Williams, R.}, \bibinfo{author}{Weinstein, A.} \&
  \bibinfo{author}{Stephens, B.}
\newblock \bibinfo{title}{{The LIGO Open Science Center}}.
\newblock \emph{\bibinfo{journal}{J. Phys. Conf. Ser.}}
  \textbf{\bibinfo{volume}{610}}, \bibinfo{pages}{012021}
  (\bibinfo{year}{2015}).

\bibitem{Su:1994gu}
\bibinfo{author}{Su, Y.} \emph{et~al.}
\newblock \bibinfo{title}{{New tests of the universality of free fall}}.
\newblock \emph{\bibinfo{journal}{Phys. Rev.}} \textbf{\bibinfo{volume}{D50}},
  \bibinfo{pages}{3614--3636} (\bibinfo{year}{1994}).

\bibitem{Schlamminger:2007ht}
\bibinfo{author}{Schlamminger, S.}, \bibinfo{author}{Choi, K.~Y.},
  \bibinfo{author}{Wagner, T.~A.}, \bibinfo{author}{Gundlach, J.~H.} \&
  \bibinfo{author}{Adelberger, E.~G.}
\newblock \bibinfo{title}{{Test of the equivalence principle using a rotating
  torsion balance}}.
\newblock \emph{\bibinfo{journal}{Phys. Rev. Lett.}}
  \textbf{\bibinfo{volume}{100}}, \bibinfo{pages}{041101}
  (\bibinfo{year}{2008}).

\bibitem{Feldman:1997qc}
\bibinfo{author}{Feldman, G.~J.} \& \bibinfo{author}{Cousins, R.~D.}
\newblock \bibinfo{title}{{A Unified approach to the classical statistical
  analysis of small signals}}.
\newblock \emph{\bibinfo{journal}{Phys. Rev.}} \textbf{\bibinfo{volume}{D57}},
  \bibinfo{pages}{3873--3889} (\bibinfo{year}{1998}).

\bibitem{LIGOScientific:2019vic}
\bibinfo{author}{Abbott, B.~P.} \emph{et~al.}
\newblock \bibinfo{title}{{Search for the isotropic stochastic background using
  data from Advanced LIGO’s second observing run}}.
\newblock \emph{\bibinfo{journal}{Phys. Rev.}} \textbf{\bibinfo{volume}{D100}},
  \bibinfo{pages}{061101} (\bibinfo{year}{2019}).

\bibitem{Foster:2017hbq}
\bibinfo{author}{Foster, J.~W.}, \bibinfo{author}{Rodd, N.~L.} \&
  \bibinfo{author}{Safdi, B.~R.}
\newblock \bibinfo{title}{{Revealing the Dark Matter Halo with Axion Direct
  Detection}}.
\newblock \emph{\bibinfo{journal}{Phys. Rev.}} \textbf{\bibinfo{volume}{D97}},
  \bibinfo{pages}{123006} (\bibinfo{year}{2018}).

\bibitem{Smith:2006ym}
\bibinfo{author}{Smith, M.~C.} \emph{et~al.}
\newblock \bibinfo{title}{{The RAVE Survey: Constraining the Local Galactic
  Escape Speed}}.
\newblock \emph{\bibinfo{journal}{Mon. Not. Roy. Astron. Soc.}}
  \textbf{\bibinfo{volume}{379}}, \bibinfo{pages}{755--772}
  (\bibinfo{year}{2007}).

\bibitem{LAL}
\bibinfo{author}{{LIGO Scientific Collaboration}}.
\newblock \bibinfo{title}{{LIGO} {A}lgorithm {L}ibrary - {LALS}uite}.
\newblock \bibinfo{howpublished}{{Free software} ({GPL}).} 
Publisher/repository https://doi.org/10.7935/GT1W-FZ16
  (\bibinfo{year}{2018}).

\end{thebibliography}
\end{document}